\documentclass[twocolumn,amsmath,aps,fleqn]{revtex4}
\usepackage{graphicx,amssymb,upgreek}
\begin{document}
%My commands
\newcommand{\be}{\begin{equation}}
\newcommand{\ee}{\end{equation}}
\newcommand{\bq}{\begin{eqnarray}}
\newcommand{\eq}{\end{eqnarray}}
\newcommand{\bsq}{\begin{subequations}}
\newcommand{\esq}{\end{subequations}}
\newcommand{\bc}{\begin{center}}
\newcommand{\ec}{\end{center}}
\newcommand {\R}{{\mathcal R}}
\newcommand{\al}{\alpha}
\newcommand\lsim{\mathrel{\rlap{\lower4pt\hbox{\hskip1pt$\sim$}}
    \raise1pt\hbox{$<$}}}
\newcommand\gsim{\mathrel{\rlap{\lower4pt\hbox{\hskip1pt$\sim$}}
    \raise1pt\hbox{$>$}}}

\title{Phantom Domain Walls}

\author{P. P. Avelino}
\email[Electronic address: ]{pedro.avelino@astro.up.pt}
\affiliation{Instituto de Astrof\'{\i}sica e Ci\^encias do Espa{\c c}o, Universidade do Porto, CAUP, Rua das Estrelas, PT4150-762 Porto, Portugal}
\affiliation{Centro de Astrof\'{\i}sica da Universidade do Porto, Rua das Estrelas, PT4150-762 Porto, Portugal}
\affiliation{Departamento de F\'{\i}sica e Astronomia, Faculdade de Ci\^encias, Universidade do Porto, Rua do Campo Alegre 687, PT4169-007 Porto, Portugal}

\author{V. M. C. Ferreira}
\email[Electronic address: ]{vasco.ferreira@astro.up.pt}
\affiliation{Instituto de Astrof\'{\i}sica e Ci\^encias do Espa{\c c}o, Universidade do Porto, CAUP, Rua das Estrelas, PT4150-762 Porto, Portugal}
\affiliation{Centro de Astrof\'{\i}sica da Universidade do Porto, Rua das Estrelas, PT4150-762 Porto, Portugal}
\affiliation{Departamento de F\'{\i}sica e Astronomia, Faculdade de Ci\^encias, Universidade do Porto, Rua do Campo Alegre 687, PT4169-007 Porto, Portugal}

\author{J. Menezes}
\email[Electronic address: ]{jmenezes@ect.ufrn.br}
\affiliation{Escola de Ci\^encias e Tecnologia, Universidade Federal do Rio Grande do Norte\\
Caixa Postal 1524, 59072-970, Natal, RN, Brazil}
\affiliation{Instituto de Astrof\'{\i}sica e Ci\^encias do Espa{\c c}o, Universidade do Porto, CAUP, Rua das Estrelas, PT4150-762 Porto, Portugal}
\affiliation{Centro de Astrof\'{\i}sica da Universidade do Porto, Rua das Estrelas, PT4150-762 Porto, Portugal}

\author{L. Sousa}
\email[Electronic address: ]{Lara.Sousa@astro.up.pt}
\affiliation{Instituto de Astrof\'{\i}sica e Ci\^encias do Espa{\c c}o, Universidade do Porto, CAUP, Rua das Estrelas, PT4150-762 Porto, Portugal}
\affiliation{Centro de Astrof\'{\i}sica da Universidade do Porto, Rua das Estrelas, PT4150-762 Porto, Portugal}

\date{\today}
\begin{abstract}

We consider a model with two real scalar fields which admits phantom domain wall solutions. We investigate the structure and evolution of these phantom domain walls in an expanding homogeneous and isotropic universe. In particular, we show that the increase of the tension of the domain walls with cosmic time, associated to the evolution of the phantom scalar field, is responsible for an additional damping term in their equations of motion. We describe the macroscopic dynamics of phantom domain walls, showing that extended phantom defects whose tension varies on a cosmological timescale cannot be the dark energy.

\end{abstract}
%\pacs{}
%\keywords{Cosmology; Dark energy}
\maketitle 

\section{\label{intr}Introduction}

Over the past years, high precision cosmological observations have been providing overwhelming evidence that the expansion of the Universe is currently accelerating (see, e.g., \cite{Suzuki:2011hu,Anderson:2012sa,Ade:2015xua}). In the standard cosmological model, this acceleration is attributed to a tiny cosmological constant which became the dominant energy component of the Universe in recent times. Despite its simplicity, a satisfactory explanation for the extremely small energy density associated to the cosmological constant is still missing. Thus, dynamical Dark Energy (DE) and modified gravity models may play a fundamental role in sourcing the acceleration of the Universe, not only at early but also at late cosmological times \cite{Copeland:2006wr,Frieman:2008sn,Caldwell:2009ix,Li:2011sd,Avelino:2016lpj}. Current observational data is perfectly consistent with dynamical DE and does not exclude the possibility that most of the energy content of our Universe might be phantom energy, as long as the value of its equation-of-state parameter is smaller than but sufficiently close to -1 \cite{Ade:2015xua}.

Topological defects, such as cosmic strings and domain walls, may leave behind a large number of interesting astrophysical and cosmological signatures. In \cite{Bucher:1998mh} it was first suggested that a domain wall network, if frozen in comoving coordinates, could be responsible for the recent acceleration of the Universe (see also \cite{Carter:2004dk,Battye:2005hw,Battye:2005ik,Carter:2006cf}). However, the possibility of a significant contribution of featureless domain walls --- defined as domain walls whose physical velocity is always perpendicular to the wall --- to the dark energy budget has since been ruled out both dynamically and observationally (the same also applies, even more strongly, in the case of line-like defects such as cosmic strings or point-like defects such as monopoles). Recently, in \cite{Dzhunushaliev:2016xdt}, compact and extended non-standard gravitating defect static solutions supported by phantom fields have been investigated, including phantom balls, strings and walls. Except for domain walls, all these solutions were shown to exhibit phantom behaviour.

In this paper we shall investigate the structure and dynamics of phantom domain walls. These are characterized by an increasing tension with cosmic time, caused by the evolution of a phantom scalar field in an expanding homogeneous and isotropic universe. In Sec. \ref{sec2}, we start by presenting a simple model with two real scalar fields which admits phantom domain wall solutions. In Sec. \ref{sec3}, we investigate the properties of static planar phantom domain wall solutions in Minkowski space. In Sec. \ref{sec4}, we extend the results of Sec. \ref{sec3} to homogeneous and isotropic Friedmann-Lema{\^ \i}tre-Robertson-Walker (FLRW) Universes, considering specific parameterizations, for definiteness. In Sec. \ref{sec5}, we compute the effect of the dynamics of the phantom field $\psi$ on the evolution of phantom domain walls. The potential role of extended phantom defects as dark energy candidates is then  discussed in Sec. \ref{sec6}. We conclude in Sec. \ref{conc}.

Throughout this paper we use units such that $c=1$, where $c$ is the value of the speed of light in vacuum.

\section{The Model \label{sec2}}

Consider the Lagrangian
\be
\mathcal L = \lambda(\psi, Y) {\mathcal V}(\phi)+f(X)\,, \label{lagrangian}
\ee
where $\phi$ and $\psi$ are real scalar fields,
\bq
X &=&  \frac12  \nabla^\mu \phi \nabla_\mu \phi\,,\\
Y &=&  \frac12  \nabla^\mu \psi \nabla_\mu \psi\,,
\eq
are their kinetic terms, $\nabla_\mu$ represents a covariant derivative with respect to the coordinate $x^\mu$, $\nabla^\mu = g^{\mu \nu}  \nabla_\nu$, and $g^{\mu \nu}$ are the components of the inverse metric tensor. Here, $\lambda(\psi, Y) < 0$ is a real function of $\psi$ and $Y$ that represents the proper pressure associated to the phantom field $\psi$, and ${\mathcal V}(\phi) \ge 0$ is a $Z_2$ symmetric potential, with two degenerate minima, that admits domain wall solutions.

The equations of motion of the scalar fields $\phi$ and $\psi$ are given, respectively, by the Euler-Lagrange equations
\bq
0 &=& - \frac{\partial {\mathcal L }}{\partial  \psi} + \nabla_\mu \left[ \frac{\partial {\mathcal L }}{\partial (\nabla_\mu \psi)}\right]\,, \label{eqofmpsi}\\
0 &=& - \frac{\partial {\mathcal L }}{\partial  \phi} + \nabla_\mu \left[ \frac{\partial {\mathcal L }}{\partial (\nabla_\mu \phi)}\right] \label{eqofmphi}\,,
\eq
or, equivalently, by
\bq
{\mathcal G}^{\mu\nu}   \nabla_\mu \nabla_\nu \psi&=& \lambda_{,\psi}-2Y \lambda_{,Y\psi}-\nonumber\\ 
& &\nabla^\gamma \psi \nabla_\gamma \phi (\ln {\mathcal V})_{,\phi} \lambda_{,Y}\,,\\
G^{\mu\nu}   \nabla_\mu \nabla_\nu \phi&=& \lambda {\mathcal V}_{,\phi}\,,
\eq
where a comma represents a partial derivative, and
\bq
{\mathcal G}^{\mu \nu}&=&{\mathcal \lambda }_{,Y} g^{\mu \nu} + {\mathcal \lambda }_{,YY}  \nabla^\mu \psi \nabla^\nu \psi\,,\\
G^{\mu \nu}&=& f_{,X} g^{\mu \nu} + f_{,XX}  \nabla^\mu \phi \nabla^\nu \phi\,.
\eq
The components of the energy-momentum tensor are
\be
T^{\mu \nu} = f_{,X} \nabla^\mu \phi \nabla^\nu \phi + {\mathcal V} \lambda_{,Y}  \nabla^\mu \psi \nabla^\nu \psi  - g^{\mu \nu} \mathcal L\,.
\ee

\section{Phantom Domain walls: Minkowski space \label{sec3}}

In this section we shall study static planar phantom domain wall solutions in Minkowski space. In this case, the line element may be written as
\be
ds^2 =  dt^2-d {\bf r} \cdot d {\bf r}\,,
\ee
where $t$ is the physical time and ${\bf r}=(x,y,z)$ are (spatial) cartesian coordinates. Consider the ansatz
\bq
\phi&=&\phi(z)\,,\\
\psi&=&\psi(t)\,,
\eq
so that $X=-{\phi'}^2/2$, $Y={\dot \psi}^2/2$ (a dot and a prime represent a derivative with respect to $t$ and $z$, respectively). Let us also define the equation-of-state parameter of the phantom field $\psi$ as
\be
w_\psi = \frac{p_\psi}{\rho_\psi}=\frac{\lambda}{\lambda_{,Y} {\dot \psi}^2 - \lambda}=\frac{\lambda}{2 \lambda_{,Y}Y - \lambda}\,,
\ee
where $\rho_\psi=2 \lambda_{,Y}Y - \lambda$ and $p_\psi=\lambda$ are the proper density and pressure associated with the phantom field $\psi$.

The components of the energy-momentum tensor are given by
\bq
\rho &=& {T^0}_0 = -{\mathcal L}+ 2 {\mathcal V} Y \lambda_{,Y}=(2 Y \lambda_{,Y}-\lambda){\mathcal V} -f \,, \label{rho}\\
p_\parallel  &=& -{T^x}_x = -{T^y}_y = {\mathcal L} = \lambda {\mathcal V} + f  =\nonumber \\
&=&  -\rho+2 {\mathcal V} \lambda_{,Y} Y   \label{ppar}\,,\\
p_\perp &=&  -{T^z}_z ={\mathcal L}- 2X f_{,X}=\lambda {\mathcal V} +f - 2 X f_{,X}=0\,,  \label{pperp}
\eq
where the last equality in Eq. (\ref{pperp}) is a consequence of energy-momentum conservation (${T^{zz}}_{,z}=0$ in Minkowski space) and of the assumption that $T^{zz} (z= \pm \infty)=0$.

It follows from Eq. (\ref{pperp}) that $2Xf_{,X}-f=\lambda {\mathcal V}$. For definiteness, let us consider that $f(X)$ can be effectively described by $f(X)=X |X|^{\alpha-1}$, with $\alpha \ge 1$. In this case, $f=\lambda {\mathcal V}/(2 \alpha -1)$ and Eqs. (\ref{rho}) and (\ref{ppar}) imply that
\bq
\rho &=& {\mathcal V} \left(\rho_\psi - \frac{p_\psi}{2 \alpha-1}\right) \,, \label{rho1}\\
p_\parallel&=&{\mathcal V} \, p_\psi \frac{2\alpha}{2 \alpha-1}\,,
\eq
which yields
\be
w_\parallel= \frac{p_\parallel}{\rho} = \frac{2\alpha w_\psi}{2\alpha-1-w_\psi}\,. \label{wparwpsi}
\ee
Eq. (\ref{wparwpsi}) implies that $w_\parallel$ is always greater than or equal to $w_\psi$ ($w_\parallel \ge w_\psi$). If $w_\psi=-1$ then $w_\parallel=w_\psi=-1$. On the other hand, if $w_\psi \ll -(2\alpha-1)$ then $w_\parallel \sim -2 \alpha$ ($w_\parallel \to-2 \alpha$ for $w_\psi \to -\infty$).

The relation between the domain wall energy per unit area and tension, defined respectively by
\bq
\sigma &\equiv & \int \rho \, dz \,, \\
{\mathcal T} & \equiv & \int p_\parallel dz \,,
\eq
is
\be
\frac{\mathcal T}{\rho} = w_\parallel\,.
\ee
Note that, for $w_\parallel \neq -1$, the components of the energy-momentum tensor are not invariant with respect to a Lorentz boost along any direction parallel to the domain wall. Hence, the physical velocity is not necessarily perpendicular to the wall and, consequently, phantom domain walls cannot be considered to be featureless.

\section{Phantom Domain Walls: FLRW background \label{sec4}}

Let us now consider a FLRW universe whose line element is given by
\be
ds^2 =  dt^2-a(t)^2 d {\bf q} \cdot d {\bf q}\,, \label{frw}
\ee
where $t$ is the physical time, $a(t)$ is the cosmological scale factor,  and ${\bf q}=(q_x,q_y,q_z)$ are comoving cartesian coordinates. For definiteness, let us assume that 
\bq
f(X)&=&X\,,\\
\lambda(\psi,Y)&=&-Y-U(\psi)\,,
\eq
with $U>0$,
and 
\bq 
{\mathcal V}(\phi)&=&V(\phi)+V_*\,,\\
V(\phi)&=&V_0\left(\frac{\phi^2}{\phi_0^2}-1\right)^2\,,\\
V_*&=&{\rm const} \ge 0\,.
\eq
In this case, the equations of motion for the scalar fields $\psi$ and $\phi$ (Eqs. (\ref{eqofmpsi}) and (\ref{eqofmphi})) yield
\bq
\ddot \psi +3 H {\dot \psi} & - & \nabla^2 \psi  =  \frac{dU}{d\psi}- \frac{d \ln {\mathcal V}}{d\phi}\left({\dot \phi} {\dot \psi} - \nabla \phi \cdot \nabla \psi \right)  \nonumber \\ 
&=& \frac{dU}{d\psi}- \frac{1}{V+V_*}\frac{d V}{d\phi}\left({\dot \phi} {\dot \psi} - \nabla \phi \cdot \nabla \psi \right)
\label{boxpsi1}\\
{\ddot \phi} +3 H {\dot \phi} & - & \nabla^2 \phi  = - |\lambda| \frac{dV}{d\phi}\,, \label{boxphi1}
\eq
where dots represent derivatives with respect to the physical time $t$, $\nabla \equiv \nabla_{\bf q}/a$, and $\nabla^2 \equiv \nabla^2_{\bf q}/a^2$.

In the $V_* \to +\infty$ limit, the last term of Eq. (\ref{boxpsi1}) is very small and can be neglected. Hence, in this case, Eq. (\ref{boxpsi1}) becomes
\be
\ddot \psi +3 H {\dot \psi} = \frac{dU}{d\psi} \label{boxpsi2}\,.
\ee
In this limit, the evolution of $\psi$ is only coupled to the evolution of $\phi$ through the source term on the right hand side of Eq. (\ref{boxphi1}). Throughout this paper we shall assume that $V_*$ is sufficiently large for it to be a good approximation to consider the equations of motion for the scalar fields $\phi$ and $\psi$ in the $V_* \to +\infty$ limit. In this case, $\psi$ and $\lambda$ may be assumed to be homogeneous in the comoving frame and the topological stability of the domain walls is guaranteed (assuming that the evolution of $\lambda$ is sufficiently slow). The study of the dynamical relevance of the last term in Eq. (\ref{boxpsi1}) shall be left for future work.

Let us start by considering the case in which $H=0$ and $\lambda = {\rm const}$. In this case, Eq. (\ref{boxphi1}) admits planar static domain wall solutions of the form
\be
\phi=\pm \phi_0 \tanh \left(\frac{z}{\delta}\right)\,,
\ee
where 
\be
\delta \sim \phi_0 (|\lambda| V_0)^{-1/2}\,,
\ee
is the thickness of the domain wall and $z=a q_z$. 

The inhomogeneous components of the energy-momentum tensor associated to the planar wall are given approximately by
\bq
\rho &=& {T^0}_0 = -V{\dot \psi}^2 +  V\left(\frac{{\dot \psi}^2}{2}+U\right)+\frac{\phi'^2}{2}\,,\\
p_\parallel  &=& -{T^x}_x = -{T^y}_y =  -V\left(\frac{{\dot \psi}^2}{2}+U\right)-\frac{\phi'^2}{2} \nonumber \\
&=& -\rho -V{\dot \psi}^2 \,,\\
p_\perp &=&  -{T^z}_z =- V\left(\frac{{\dot \psi}^2}{2}+U\right)+\frac{\phi'^2}{2} = 0\label{pperp1}\,,
\eq
where $p_\perp$ vanishes as a result of energy-momentum conservation as discussed in the previous section.

By using Eq. (\ref{pperp1}), one finds that

\bq
\rho & = & 2UV > 0\,,\\
p_\parallel & = & -V\left(2U+{\dot \psi}^2\right)\,,
\eq
so that $w_\perp=p_\perp/\rho=0$, and
\be
w_\parallel= \frac{p_\parallel}{\rho} = -1-\frac{{\dot \psi}^2}{2U}=\frac{2 w_\psi}{1-w_{\psi}}\,, \label{wparwpsi1}
\ee
where $\rho_\psi=-{\dot \psi}^2/2+U$ and $p_\psi=\lambda=-{\dot \psi}^2/2-U$. Eq. (\ref{wparwpsi1}) is equivalent to Eq. (\ref{wparwpsi}) for $\alpha=1$.

Let us now consider the case of a frozen phantom domain wall --- so that the region with $\phi=0$ (or, equivalently, maximum $V(t)$) does not move --- with $\lambda= \lambda(t)$ in a FLRW universe with $H \neq 0$, and introduce three characteristic timescales defined by
\bq
{\Delta t}_{\delta}&=& \delta \sim \phi_0 (|\lambda| V_0)^{-1/2}\,,\\
{\Delta t}_\lambda &=& \left| \frac{\dot \lambda}{\lambda}\right|^{-1} \sim \left|\frac{\dot \delta}{\delta}\right|^{-1}\,,\\
{\Delta t}_H &=& H^{-1}\,,
\eq
which represent, respectively, the time necessary for light to travel across a domain wall of thickness $\delta$, the characteristic time associated to variations of $\delta$ and $\lambda$, and the Hubble time. In the most interesting situation, in which ${\Delta t}_\delta \ll {\Delta t}_\lambda$ and ${\Delta t}_\delta \ll {\Delta t}_H$, the results obtained previously for $H=0$ and $\lambda = ª\rm const$ are still approximately valid, except that now both $\psi$ and $\lambda$ are assumed to be generic functions of the physical time alone. 

Although the condition $d |\lambda| / dt > 0$ would be automatically satisfied if  $w_\psi = {\rm const} < -1$, it might not be true in general. Nevertheless, throughout this paper we shall assume that the time dependence of $w_\psi$ is such that  $d |\lambda| / dt > 0$ is always verified. In this case, the thickness of domain walls is thus affected by the evolution of the phantom field and decreases, proportionally to $|\lambda|^{-1/2}$, as the expansion of the background causes $|\lambda|$ to increase. The opposite happens to the domain wall tension which increases proportionally to $|\lambda|$.

\section{Phantom domain wall dynamics \label{sec5}}

In this section, we shall use the method devised in \cite{Sousa:2010zza} to extract the dynamics of phantom domain walls from Eq. (\ref{boxphi1}). Let us start by performing a change of variables in Eq. (\ref{boxphi1}) from $(t,\bf{q})$ into a new coordinate set $(\tau,\bf{u})$ defined by
\be
\frac{\partial}{\partial \tau}=\frac{1}{|\lambda|^{1/2}}\frac{\partial}{\partial t}\,,\quad\mbox{and}\quad \mathbf{u}= a |\lambda|^{1/2}\mathbf{q}\,.
\ee

In this case, Eq. (\ref{boxphi1}) yields
\be
\frac{\partial ^2 \phi}{\partial \tau ^2}+\left(3H_\tau+\frac{1}{2}\frac{d\ln{\left|\lambda\right|}}{d\tau}\right)\frac{\partial \phi}{\partial \tau}-\nabla_{\bf{u}}^2\phi=-\frac{dV}{d\phi}\,,
\ee
where $\nabla_{\bf{u}}^2=|\lambda|^{-1}\nabla^2$ and $H_\tau=|\lambda|^{-1/2}H$.

In Minkowski spacetime, a planar static domain wall solution oriented perpendicularly to the $z$ direction is given by $\phi=\phi_s(l)$ and satisfying
\be
\frac{d^2 \phi_s}{d l^2}=\frac{dV}{d\phi}\label{staticwall}\,,
\ee
where $l=u_z$ (we took ${\bf{u}}=(u_x,u_y,u_z)$). If the domain wall is boosted with a velocity $v$ along the positive $z$ direction, the planar solution still satisfies Eq. (\ref{staticwall}), but now we have $l=\gamma(u_z-v\tau)$.

Let us now consider the more general case of a curved domain wall section in a $3+1$-dimensional FLRW universe and assume that it is locally flat (or equivalently that its thickness is very small when compared to its curvature radii). In this case, we may choose a new set of $\mathbf{u}$ coordinates such that the domain wall is locally defined by $u_z={\rm const}$ and it moves along the positive $u_z$ direction. Moreover, we shall use gauge freedom to choose a coordinate $u_z$ which measures the arc-length along the direction perpendicular to the wall. Once again, the domain wall solution with still be given by $\phi=\phi_s(l)$ (satisfying Eq. (\ref{staticwall}), with $l=\gamma(u_z-v\tau)$). We then have that
\bq
\frac{\partial\phi_s}{\partial\tau}  &=&  -\gamma v\frac{d\phi_s}{dl}\,,\\
\frac{\partial^2\phi_s}{\partial \tau^2} &=& \left(\gamma v\right)^2\frac{d^2\phi_s}{dl^2}-\frac{d \left(\gamma v\right)}{d\tau}\frac{d\phi_s}{dl}\,,\\
\frac{\partial\phi_s}{\partial u}  &=&  \gamma\frac{d\phi_s}{dl}\,,\quad \quad\frac{\partial^2 \phi_s}{\partial^2 u}=\gamma^2\frac{d^2\phi_s}{dl^2}\,.
\eq

Moreover, it was shown in \cite{Sousa:2010zza} that
\be
\nabla_{\bf{u}}^2\phi_s=-\gamma\kappa_{\bf u}\frac{d\phi_s}{dl}+\gamma^2\frac{d^2\phi_s}{d l^2}\,,
\ee
where $\kappa_{\bf u}=|\lambda|^{-1/2}\kappa$ is the extrinsic curvature measured in the non-physical $\bf u$ coordinates, and $\kappa$ is the physical curvature.

We then have that
\be
-\frac{d^2\phi_s}{dl^2}+\mathcal{F}\frac{d\phi_s}{dl}=-\frac{dV}{d\phi_s}\,,
\ee
with
\be
\mathcal{F}=-\frac{d}{d\tau}\left(\gamma v\right)-\gamma v\left(3H_\tau+\frac{1}{2}\frac{d\ln{\left|\lambda\right|}}{d\tau}\right)+\gamma \kappa_{\bf u}\,.
\ee
Since $\phi_s(l)$ must necessarily satisfy Eq. (\ref{staticwall}), we should then have that $\mathcal{F}=0$ or, equivalently, that
\be
\frac{dv}{dt}=\left(1-v^2\right)\left[\kappa-v\left(3H+\frac{1}{2}\left|\frac{\dot \lambda}{\lambda}\right|\right)\right]\label{eomv}\,.
\ee
Therefore, phantom domain walls feel an additional damping effect caused by the increase of their tension with time. Also, since we are assuming that the scalar field $\psi$ is homogeneous in the comoving frame, the domain wall velocity, in this reference frame, is determined by the evolution of the scalar field $\phi$ and is, therefore, perpendicular to the wall. Notice that Eq. (\ref{eomv}) is equivalent to the evolution equation for the velocity of a domain wall with varying tension introduced in \cite{Avelino:2015kdn}.

\section{Phantom Domain Walls and Dark Energy\label{sec6}}

The dynamics of networks of topological defects of arbitrary dimensionality may be described statistically on sufficiently large scales  by resorting to a Velocity-dependent One-Scale (VOS) model \cite{Sousa:2011ew,Sousa:2011iu}. This model describes the macroscopic evolution of topological defects by following two variables: the characteristic length $L$ --- defined, in the case of domain walls, as

\be
{\bar \rho}=\frac{\sigma}{L}\,,
\ee
where $\sigma$ is the surface energy density and ${\bar \rho}$ is the average domain wall energy density --- and the Root-Mean-Squared (RMS) velocity ${\bar v}$. 

For phantom domain walls, an evolution equation for the RMS velocity of the network may be obtained by averaging Eq. (\ref{eomv}) for the whole network. This yields
\be
\frac{d{\bar v}}{dt}= \left(1-{\bar v}^2\right)\left(\frac{k}{L}-\frac{\bar v}{\ell_d}\right)\,,\label{vosv}
\ee
where we have defined the damping lengthscale as
\be
\ell_d^{-1}=3H+\frac{1}{2}\left|\frac{\dot \lambda}{\lambda}\right|\,.
\label{elld}
\ee
Here, $k({\bar v})$ is the averaged momentum parameter --- rigorously defined in \cite{Avelino:2011ev,Sousa:2011ew} --- which describes the conversion of rest mass energy into kinetic energy (and vice-versa) by the network, thus describing the acceleration caused by domain wall curvature. Note that this equation assumes the same form as the evolution equation for $\bar v$ for standard domain walls  \cite{Avelino:2011ev,Sousa:2011ew}, but with a modified damping lengthscale. $\ell_d$ here includes the damping effect caused by the decrease of domain wall thickness associated to the non-minimal coupling to the phantom field.

Since phantom domain walls are able to provide a negative average pressure, the question of whether they can contribute to the dark energy budget naturally emerges. Current Cosmic Microwave Background (CMB) constraints restrain the fractional density of domain walls with a characteristic length comparable to the cosmological horizon to be
\be
\Omega_\sigma=\frac{\bar \rho}{\rho_c}<10^{-5}\,,
\ee
where $\rho_c$ is the critical density of the universe (see Ref. \cite{Sousa:2015cqa} for a detailed characterization of the CMB constraints on standard domain wall networks). If phantom domain walls are to contribute significantly to the dark energy budget, one would need $\Omega_\sigma\sim\mathcal{O}(1)$ and, thus, a characteristic length significantly smaller than the cosmological horizon.

As discussed in Ref. \cite{Avelino:2008ve,Sousa:2009is}, the amplitude of the CMB temperature fluctuations generated by domain walls are (conservatively) constrained around the present time to be smaller than $10^{-5}$ down to scales of the order of $L_V=H_0^{-1}/100$ (otherwise they would generate strong signatures on the CMB). This means that the fractional density fluctuations associated with domain walls on a physical scale $L_V$ (much larger than $L$),
\be
\updelta \equiv \frac{\delta {\bar \rho}}{\rho_c}\sim \frac{\Omega_\sigma}{\sqrt{N}}\,,
\ee
--- where $N\sim (L_V/L)^3$ is the number of domain walls on a volume $V=L_V^3$ and $\delta{\bar \rho}$ are the RMS fluctuations on the domain wall energy density on the $L_V$ physical scale --- should be
\be
\updelta \sim 10^3\Omega_{\sigma 0}\left(H_0 L_0\right)^{3/2} \lesssim 10^{-5}\,.
\ee

We should then have that
\be
H_0L_0\lesssim 10^{-5} \Omega_{\sigma 0}^{-2/3}\,,
\ee
and consequently
\be
{\bar v}_0\lesssim 10^{-5} \Omega_{\sigma 0}^{-2/3}\,,
\ee
where we have used the fact that one should expect ${\bar v}_0\lesssim H_0 L_0$, unless there is an abrupt increase of  the velocity of domain walls near the present time. Therefore, if phantom domain walls are to significantly contribute to the dark energy budget, their RMS velocity should be extremely small and the network should therefore be frustrated (or frozen in comoving coordinates). The dynamics of frustrated domain wall networks have been thoroughly studied in the literature \cite{PinaAvelino:2006ia,Avelino:2008ve,Sousa:2009is,Sousa:2011iu} and all studies indicate that such  networks cannot realistically be (or significantly contribute to) dark energy. The frustration of domain wall networks can be achieved with additional damping mechanisms, however the energy necessary to decelerate the walls is so large that these mechanisms would make a much larger contribution than the walls themselves to the energy budget of the universe. As a consequence, in order not to spoil current observational data, domain walls would need to have an energy density that is significantly smaller than the critical density, thus contributing negligibly to dark energy. As we shall see here, similar arguments apply in the case of phantom domain walls.

The curvature term in Eq. (\ref{vosv}) causes the domain walls to accelerate and, hence, it has to be suppressed in order for frustration to occur. It was shown that, for standard domain wall networks, $k\sim 1$ and that Hubble damping is insufficient to freeze domain walls in this case  \cite{PinaAvelino:2006ia}. Although more complex networks with junctions exhibit smaller values of $k$, in this case $k$ is still of order unity. As matter of fact, one would not expect $k\ll 1$ in any realistic domain wall network for causality reasons. Therefore, the only way in which frustration can be achieved for phantom domain walls is if the damping caused by the variations of domain wall thickness is strong enough to counteract the effects of curvature:
\be
\frac{k}{L_0} \lesssim\left|\frac{\dot{\lambda}}{\lambda}\right|{\bar v}_0\,,
\ee
or equivalently if
\be
\frac{1}{H_0}\left|\frac{\dot{\lambda}}{\lambda}\right|\gtrsim \frac{1}{{\bar v}_0 H_0 L_0}\,,
\ee
where we have used the fact that $k\sim 1$. We should then have that
\be
\frac{1}{H_0}\left|\frac{\dot{\lambda}}{\lambda}\right|\gtrsim 10^{10}\Omega_{\sigma 0}^{4/3}\,.
\ee
Hence, if phantom domain walls are to contribute significantly to the dark energy budget, with $\Omega_{\sigma 0}\sim 1$, the characteristic timescale associated with the variation of $\lambda$ would have to be much smaller than one Hubble time (that is, ${\Delta t}_\lambda \ll {\Delta t}_H$) at recent times. However, since the expansion of the universe plays a crucial role in feeding the time variation of $\lambda(t)$, one would not expect this to be case. Note that similar constraints apply to other extended phantom defects, such as phantom cosmic strings, and therefore --- despite also being able to provide an average negative pressure --- these are also not expected to contribute significantly to the dark energy budget.

\section{\label{conc} Conclusions}

In this paper we introduced a simple model with two real scalar fields which admits phantom domain wall configurations. We computed the corresponding solutions in Minkowski and FLRW spacetimes, showing that in an expanding FLRW universe an increasing tension, associated to the evolution of a phantom scalar field, gives rise to an additional damping term in their equations of motion. We have further shown  that extended phantom defects, such as phantom domain walls, whose tension varies on a cosmological timescale cannot be the dark energy.

%%%%%%%%%%%%%%%%%%%%%%%%%%%%%%%%%%%%%%%%%%%%%%%%%%%%%
\begin{acknowledgments}

V.M.C. Ferreira was supported by FCT through national funds and by FEDER through COMPETE2020 (ref: POCI-01-0145-FEDER-007672). L.S. is supported by Funda{\c c}\~ao para a Ci\^encia e a Tecnologia (FCT, Portugal) and by the European Social Fund (POPH/FSE) through the grant SFRH/BPD/76324/2011. Funding of this work was also provided by the FCT grant UID/FIS/04434/2013.

\end{acknowledgments}

%%%%%%%%%%%%%%%%%%%%%%%%%%%%%%%%%%%%%%%%%%%%%%%%%%%%%%%%%%

\bibliography{Phantom_defects}

\begin{thebibliography}{23}
\expandafter\ifx\csname natexlab\endcsname\relax\def\natexlab#1{#1}\fi
\expandafter\ifx\csname bibnamefont\endcsname\relax
  \def\bibnamefont#1{#1}\fi
\expandafter\ifx\csname bibfnamefont\endcsname\relax
  \def\bibfnamefont#1{#1}\fi
\expandafter\ifx\csname citenamefont\endcsname\relax
  \def\citenamefont#1{#1}\fi
\expandafter\ifx\csname url\endcsname\relax
  \def\url#1{\texttt{#1}}\fi
\expandafter\ifx\csname urlprefix\endcsname\relax\def\urlprefix{URL }\fi
\providecommand{\bibinfo}[2]{#2}
\providecommand{\eprint}[2][]{\url{#2}}

\bibitem[{\citenamefont{Suzuki et~al.}(2012)\citenamefont{Suzuki, Rubin,
  Lidman, Aldering, Amanullah et~al.}}]{Suzuki:2011hu}
\bibinfo{author}{\bibfnamefont{N.}~\bibnamefont{Suzuki}},
  \bibinfo{author}{\bibfnamefont{D.}~\bibnamefont{Rubin}},
  \bibinfo{author}{\bibfnamefont{C.}~\bibnamefont{Lidman}},
  \bibinfo{author}{\bibfnamefont{G.}~\bibnamefont{Aldering}},
  \bibinfo{author}{\bibfnamefont{R.}~\bibnamefont{Amanullah}},
  \bibnamefont{et~al.}, \bibinfo{journal}{Astrophys.J.}
  \textbf{\bibinfo{volume}{746}}, \bibinfo{pages}{85} (\bibinfo{year}{2012}),
  \eprint{1105.3470}.

\bibitem[{\citenamefont{Anderson et~al.}(2013)\citenamefont{Anderson, Aubourg,
  Bailey, Bizyaev, Blanton et~al.}}]{Anderson:2012sa}
\bibinfo{author}{\bibfnamefont{L.}~\bibnamefont{Anderson}},
  \bibinfo{author}{\bibfnamefont{E.}~\bibnamefont{Aubourg}},
  \bibinfo{author}{\bibfnamefont{S.}~\bibnamefont{Bailey}},
  \bibinfo{author}{\bibfnamefont{D.}~\bibnamefont{Bizyaev}},
  \bibinfo{author}{\bibfnamefont{M.}~\bibnamefont{Blanton}},
  \bibnamefont{et~al.}, \bibinfo{journal}{Mon.Not.Roy.Astron.Soc.}
  \textbf{\bibinfo{volume}{427}}, \bibinfo{pages}{3435} (\bibinfo{year}{2013}),
  \eprint{1203.6594}.

\bibitem[{\citenamefont{Ade et~al.}(2016)}]{Ade:2015xua}
\bibinfo{author}{\bibfnamefont{P.~A.~R.} \bibnamefont{Ade}}
  \bibnamefont{et~al.} (\bibinfo{collaboration}{Planck}),
  \bibinfo{journal}{Astron. Astrophys.} \textbf{\bibinfo{volume}{594}},
  \bibinfo{pages}{A13} (\bibinfo{year}{2016}), \eprint{1502.01589}.

\bibitem[{\citenamefont{Copeland et~al.}(2006)\citenamefont{Copeland, Sami, and
  Tsujikawa}}]{Copeland:2006wr}
\bibinfo{author}{\bibfnamefont{E.~J.} \bibnamefont{Copeland}},
  \bibinfo{author}{\bibfnamefont{M.}~\bibnamefont{Sami}}, \bibnamefont{and}
  \bibinfo{author}{\bibfnamefont{S.}~\bibnamefont{Tsujikawa}},
  \bibinfo{journal}{Int. J. Mod. Phys.} \textbf{\bibinfo{volume}{D15}},
  \bibinfo{pages}{1753} (\bibinfo{year}{2006}), \eprint{hep-th/0603057}.

\bibitem[{\citenamefont{Frieman et~al.}(2008)\citenamefont{Frieman, Turner, and
  Huterer}}]{Frieman:2008sn}
\bibinfo{author}{\bibfnamefont{J.}~\bibnamefont{Frieman}},
  \bibinfo{author}{\bibfnamefont{M.}~\bibnamefont{Turner}}, \bibnamefont{and}
  \bibinfo{author}{\bibfnamefont{D.}~\bibnamefont{Huterer}},
  \bibinfo{journal}{Ann. Rev. Astron. Astrophys.}
  \textbf{\bibinfo{volume}{46}}, \bibinfo{pages}{385} (\bibinfo{year}{2008}).

\bibitem[{\citenamefont{Caldwell and Kamionkowski}(2009)}]{Caldwell:2009ix}
\bibinfo{author}{\bibfnamefont{R.~R.} \bibnamefont{Caldwell}} \bibnamefont{and}
  \bibinfo{author}{\bibfnamefont{M.}~\bibnamefont{Kamionkowski}},
  \bibinfo{journal}{Ann. Rev. Nucl. Part. Sci.} \textbf{\bibinfo{volume}{59}},
  \bibinfo{pages}{397} (\bibinfo{year}{2009}).

\bibitem[{\citenamefont{Li et~al.}(2011)\citenamefont{Li, Li, Wang, and
  Wang}}]{Li:2011sd}
\bibinfo{author}{\bibfnamefont{M.}~\bibnamefont{Li}},
  \bibinfo{author}{\bibfnamefont{X.-D.} \bibnamefont{Li}},
  \bibinfo{author}{\bibfnamefont{S.}~\bibnamefont{Wang}}, \bibnamefont{and}
  \bibinfo{author}{\bibfnamefont{Y.}~\bibnamefont{Wang}},
  \bibinfo{journal}{Commun.Theor.Phys.} \textbf{\bibinfo{volume}{56}},
  \bibinfo{pages}{525} (\bibinfo{year}{2011}).

\bibitem[{\citenamefont{Avelino et~al.}(2016)}]{Avelino:2016lpj}
\bibinfo{author}{\bibfnamefont{P.~P.} \bibnamefont{Avelino}}
  \bibnamefont{et~al.}, \bibinfo{journal}{Symmetry}
  \textbf{\bibinfo{volume}{8}}, \bibinfo{pages}{70} (\bibinfo{year}{2016}),
  \eprint{1607.02979}.

\bibitem[{\citenamefont{Bucher and Spergel}(1999)}]{Bucher:1998mh}
\bibinfo{author}{\bibfnamefont{M.}~\bibnamefont{Bucher}} \bibnamefont{and}
  \bibinfo{author}{\bibfnamefont{D.~N.} \bibnamefont{Spergel}},
  \bibinfo{journal}{Phys. Rev.} \textbf{\bibinfo{volume}{D60}},
  \bibinfo{pages}{043505} (\bibinfo{year}{1999}), \eprint{astro-ph/9812022}.

\bibitem[{\citenamefont{Carter}(2005)}]{Carter:2004dk}
\bibinfo{author}{\bibfnamefont{B.}~\bibnamefont{Carter}},
  \bibinfo{journal}{Int. J. Theor. Phys.} \textbf{\bibinfo{volume}{44}},
  \bibinfo{pages}{1729} (\bibinfo{year}{2005}), \eprint{hep-ph/0412397}.

\bibitem[{\citenamefont{Battye et~al.}(2005)\citenamefont{Battye, Carter,
  Chachoua, and Moss}}]{Battye:2005hw}
\bibinfo{author}{\bibfnamefont{R.~A.} \bibnamefont{Battye}},
  \bibinfo{author}{\bibfnamefont{B.}~\bibnamefont{Carter}},
  \bibinfo{author}{\bibfnamefont{E.}~\bibnamefont{Chachoua}}, \bibnamefont{and}
  \bibinfo{author}{\bibfnamefont{A.}~\bibnamefont{Moss}},
  \bibinfo{journal}{Phys. Rev.} \textbf{\bibinfo{volume}{D72}},
  \bibinfo{pages}{023503} (\bibinfo{year}{2005}), \eprint{hep-th/0501244}.

\bibitem[{\citenamefont{Battye et~al.}(2006)\citenamefont{Battye, Chachoua, and
  Moss}}]{Battye:2005ik}
\bibinfo{author}{\bibfnamefont{R.~A.} \bibnamefont{Battye}},
  \bibinfo{author}{\bibfnamefont{E.}~\bibnamefont{Chachoua}}, \bibnamefont{and}
  \bibinfo{author}{\bibfnamefont{A.}~\bibnamefont{Moss}},
  \bibinfo{journal}{Phys. Rev.} \textbf{\bibinfo{volume}{D73}},
  \bibinfo{pages}{123528} (\bibinfo{year}{2006}), \eprint{hep-th/0512207}.

\bibitem[{\citenamefont{Carter}(2008)}]{Carter:2006cf}
\bibinfo{author}{\bibfnamefont{B.}~\bibnamefont{Carter}},
  \bibinfo{journal}{Class. Quant. Grav.} \textbf{\bibinfo{volume}{25}},
  \bibinfo{pages}{154001} (\bibinfo{year}{2008}), \eprint{hep-ph/0605029}.

\bibitem[{\citenamefont{Dzhunushaliev et~al.}(2016)\citenamefont{Dzhunushaliev,
  Folomeev, Makhmudov, Urazalina, Singleton, and
  Scott}}]{Dzhunushaliev:2016xdt}
\bibinfo{author}{\bibfnamefont{V.}~\bibnamefont{Dzhunushaliev}},
  \bibinfo{author}{\bibfnamefont{V.}~\bibnamefont{Folomeev}},
  \bibinfo{author}{\bibfnamefont{A.}~\bibnamefont{Makhmudov}},
  \bibinfo{author}{\bibfnamefont{A.}~\bibnamefont{Urazalina}},
  \bibinfo{author}{\bibfnamefont{D.}~\bibnamefont{Singleton}},
  \bibnamefont{and} \bibinfo{author}{\bibfnamefont{J.}~\bibnamefont{Scott}},
  \bibinfo{journal}{Phys. Rev.} \textbf{\bibinfo{volume}{D94}},
  \bibinfo{pages}{024004} (\bibinfo{year}{2016}), \eprint{1606.07304}.

\bibitem[{\citenamefont{Sousa and Avelino}(2010{\natexlab{a}})}]{Sousa:2010zza}
\bibinfo{author}{\bibfnamefont{L.}~\bibnamefont{Sousa}} \bibnamefont{and}
  \bibinfo{author}{\bibfnamefont{P.~P.} \bibnamefont{Avelino}},
  \bibinfo{journal}{Phys. Rev.} \textbf{\bibinfo{volume}{D81}},
  \bibinfo{pages}{087305} (\bibinfo{year}{2010}{\natexlab{a}}),
  \eprint{1101.3350}.

\bibitem[{\citenamefont{Avelino and Sousa}(2016)}]{Avelino:2015kdn}
\bibinfo{author}{\bibfnamefont{P.~P.} \bibnamefont{Avelino}} \bibnamefont{and}
  \bibinfo{author}{\bibfnamefont{L.}~\bibnamefont{Sousa}},
  \bibinfo{journal}{Phys. Rev.} \textbf{\bibinfo{volume}{D93}},
  \bibinfo{pages}{023519} (\bibinfo{year}{2016}), \eprint{1511.00589}.

\bibitem[{\citenamefont{Sousa and Avelino}(2011{\natexlab{a}})}]{Sousa:2011iu}
\bibinfo{author}{\bibfnamefont{L.}~\bibnamefont{Sousa}} \bibnamefont{and}
  \bibinfo{author}{\bibfnamefont{P.~P.} \bibnamefont{Avelino}},
  \bibinfo{journal}{Phys. Rev.} \textbf{\bibinfo{volume}{D84}},
  \bibinfo{pages}{063502} (\bibinfo{year}{2011}{\natexlab{a}}),
  \eprint{1107.4582}.

\bibitem[{\citenamefont{Sousa and Avelino}(2011{\natexlab{b}})}]{Sousa:2011ew}
\bibinfo{author}{\bibfnamefont{L.}~\bibnamefont{Sousa}} \bibnamefont{and}
  \bibinfo{author}{\bibfnamefont{P.~P.} \bibnamefont{Avelino}},
  \bibinfo{journal}{Phys. Rev.} \textbf{\bibinfo{volume}{D83}},
  \bibinfo{pages}{103507} (\bibinfo{year}{2011}{\natexlab{b}}),
  \eprint{1103.1381}.

\bibitem[{\citenamefont{Avelino and Sousa}(2011)}]{Avelino:2011ev}
\bibinfo{author}{\bibfnamefont{P.~P.} \bibnamefont{Avelino}} \bibnamefont{and}
  \bibinfo{author}{\bibfnamefont{L.}~\bibnamefont{Sousa}},
  \bibinfo{journal}{Phys. Rev.} \textbf{\bibinfo{volume}{D83}},
  \bibinfo{pages}{043530} (\bibinfo{year}{2011}), \eprint{1101.3360}.

\bibitem[{\citenamefont{Sousa and Avelino}(2015)}]{Sousa:2015cqa}
\bibinfo{author}{\bibfnamefont{L.}~\bibnamefont{Sousa}} \bibnamefont{and}
  \bibinfo{author}{\bibfnamefont{P.~P.} \bibnamefont{Avelino}}
  (\bibinfo{year}{2015}), \eprint{1507.01064}.

\bibitem[{\citenamefont{Avelino et~al.}(2008)\citenamefont{Avelino, Martins,
  Menezes, Menezes, and Oliveira}}]{Avelino:2008ve}
\bibinfo{author}{\bibfnamefont{P.~P.} \bibnamefont{Avelino}},
  \bibinfo{author}{\bibfnamefont{C.~J. A.~P.} \bibnamefont{Martins}},
  \bibinfo{author}{\bibfnamefont{J.}~\bibnamefont{Menezes}},
  \bibinfo{author}{\bibfnamefont{R.}~\bibnamefont{Menezes}}, \bibnamefont{and}
  \bibinfo{author}{\bibfnamefont{J.~C. R.~E.} \bibnamefont{Oliveira}},
  \bibinfo{journal}{Phys. Rev.} \textbf{\bibinfo{volume}{D78}},
  \bibinfo{pages}{103508} (\bibinfo{year}{2008}), \eprint{0807.4442}.

\bibitem[{\citenamefont{Sousa and Avelino}(2010{\natexlab{b}})}]{Sousa:2009is}
\bibinfo{author}{\bibfnamefont{L.}~\bibnamefont{Sousa}} \bibnamefont{and}
  \bibinfo{author}{\bibfnamefont{P.~P.} \bibnamefont{Avelino}},
  \bibinfo{journal}{Phys. Lett.} \textbf{\bibinfo{volume}{B689}},
  \bibinfo{pages}{145} (\bibinfo{year}{2010}{\natexlab{b}}),
  \eprint{0911.3902}.

\bibitem[{\citenamefont{Avelino et~al.}(2006)\citenamefont{Avelino, Martins,
  Menezes, Menezes, and Oliveira}}]{PinaAvelino:2006ia}
\bibinfo{author}{\bibfnamefont{P.~P.} \bibnamefont{Avelino}},
  \bibinfo{author}{\bibfnamefont{C.~J. A.~P.} \bibnamefont{Martins}},
  \bibinfo{author}{\bibfnamefont{J.}~\bibnamefont{Menezes}},
  \bibinfo{author}{\bibfnamefont{R.}~\bibnamefont{Menezes}}, \bibnamefont{and}
  \bibinfo{author}{\bibfnamefont{J.~C. R.~E.} \bibnamefont{Oliveira}},
  \bibinfo{journal}{Phys. Rev.} \textbf{\bibinfo{volume}{D73}},
  \bibinfo{pages}{123519} (\bibinfo{year}{2006}), \eprint{astro-ph/0602540}.

\end{thebibliography}

\end{document}